\documentclass[%
twocolumn,
superscriptaddress,
amsmath,amssymb,
aps, physrev,
]{revtex4-2}

\usepackage{graphicx} 
\usepackage[colorlinks,allcolors=black]{hyperref} 
\usepackage[capitalise]{cleveref}
\usepackage{microtype}
\usepackage[export]{adjustbox}

\DeclareMathOperator*{\argmax}{arg\,max}

\newcommand{\vi}{\boldsymbol{v}_i}
\newcommand{\vpriv}{\boldsymbol{v}_i^{\text{priv}}}
\newcommand{\vpub}{\boldsymbol{v}_i^{\text{pub}}} 
 
\newcommand{\vcm}{\boldsymbol{v}_\text{cm}} 
\newcommand{\tmem}{t_{\text{mem}}}
\newcommand{\rd}{\text{R}_{\text{d}}}
\newcommand{\rv}{\text{R}_{\text{v}}}
\newcommand{\rs}{\text{R}_{\text{s}}}
\newcommand{\rc}{\gamma_i}
\begin{document}


\title{\textbf{
Zero-information limit of a collective olfactory search model
}}%

\author{Francesco Boccardo}
\affiliation{MaLGa, Department of Civil, Chemical and Environmental Engineering, University of Genoa, Genoa, Italy}

\author{Simone Di Marino}
\affiliation{MaLGa, Department of Mathematics (DIMA), University of Genoa, Genoa, Italy}

\author{Agnese Seminara}
\affiliation{MaLGa, Department of Civil, Chemical and Environmental Engineering, University of Genoa, Genoa, Italy}

\date{\today}

\begin{abstract}
We address the problem of how individuals can integrate efficiently their private behavior with information provided by others within a group.
To this end, we consider the model of collective search introduced in Ref.~\cite{Durve2020}, under a minimal setting with no olfactory information. 
Agents combine a private exploratory behavior and a social imitation consisting in aligning to their neighbors, and weigh the two contributions with a single ``trust" parameter that controls their relative influence.
We find that an optimal trust parameter exists even in the absence of olfactory information, as was observed in the original model. Optimality is dictated by the need to explore the minimal region of space that contains the target. An optimal trust parameter emerges from this constraint because it it tunes imitation, which induces a collective mechanism of inertia affecting the size and path of the swarm.
We predict the optimal trust parameter for cohesive groups where all agents interact with one another.
We show how optimality depends on the initialization of the agents and the unknown location of the target, in close agreement with numerical simulations.
Our results may be leveraged to optimize the design of swarm robotics or to understand information integration in organisms with decentralized nervous systems such as cephalopods.

\end{abstract}

\maketitle

\section{Introduction}

The combination of individual sensing and social communication gives rise to collective behaviors that can surpass the capabilities of single agents in both biological and artificial systems \cite{Couzin2009, Berdahl2013, Bonabeau1999, STEPHENS2019}. A key factor in such systems is how information is shared among members of the group \cite{Torney2018, Kao2014}. 
This is especially relevant in scenarios where sensory information is sparse or even entirely absent, forcing agents to rely almost entirely on one another.  
Several models have been proposed that aim at combining individual and social behavior, for example in the study of crowd dynamics \cite{Helbing2000}, migration dynamics \cite{Torney2010}, consensus decision making \cite{Miller2013} and collective navigation driven by light or smell \cite{Torney2009, Durve2020, Berdahl2013}. 

In this work, we investigate the zero-information limit of a collective search model inspired by olfactory navigation of moths, and originally introduced by Durve \textit{et al.} \cite{Durve2020}. Within a collective search for a fixed target, each agent responds to two distinct types of information: a private signal, which drives an individual exploratory behavior and a public signal based on the motion of nearby agents, which induces social imitation in the form of a Vicsek-like alignment \cite{Vicsek1995,Vicsek2012}. The relative influence of these two behavioral responses is controlled by a single tunable parameter, the trust parameter $\beta$, which determines how much weight an agent places on social alignment versus private exploration.
In the presence of odor, an optimal trust parameter $\beta^*$ was observed, where group performance was maximized. However, what are the origins of this optimality and how it may shift in response to the conditions of the search is not understood.
While the original model was developed in the presence of a turbulent olfactory signal, here we analyze its dynamics in the zero-information limit i.e. in the complete absence of odor. The zero-information limit reveals key geometric constraints on social interaction, regardless of environmental cues. 

This framework is relevant to swarms of independent agents, as well as to organisms with distributed nervous systems, such as cephalopods. For example, octopuses 
have flexible bodies whose movement is controlled locally by neural ganglia distributed along the arm, with coordination provided by the central brain \cite{GodfreySmith2016, Sumbre2006}. In this context, the trust parameter $\beta$ can be interpreted as the weight of central coordination over local control. In this case, agents receive information from all other members of the group, and this full internal communication corresponds to an infinite-range interaction between the sensors. Similar trade-offs between local autonomy and centralized control also arise in robotic swarms and distributed artificial systems \cite{Brambilla2013}.

Here we combine agent-based simulations and asymptotics and show that an optimal trust parameter exists even in the absence of odor.
A key result of our study is that the individual trajectories stretch upwind as a result of imitation amplifying intrinsic directional biases and ultimately dictating how the swarm collectively explores space. 
We develop an asymptotic theory to quantify the geometry of the area explored by a cohesive swarm where all agents interact and show that trusting others boils down to an emerging mechanism of group inertia. 
We then derive the optimal trust parameter by imposing that the swarm explores the smallest region of space containing the target. The theory is in close agreement with the numerical simulations and shows how optimality shifts with the initial conditions and the location of the target. 

Clearly, without odor the search is much slower. However, the optimal trust parameter is qualitatively consistent with what was observed in the original model in the presence of odor \cite{Durve2020}, suggesting that geometry may remain a relevant player even when sensory information is added back into the picture.  

\begin{figure}[ht]
    \centering
    \includegraphics[width=\columnwidth]{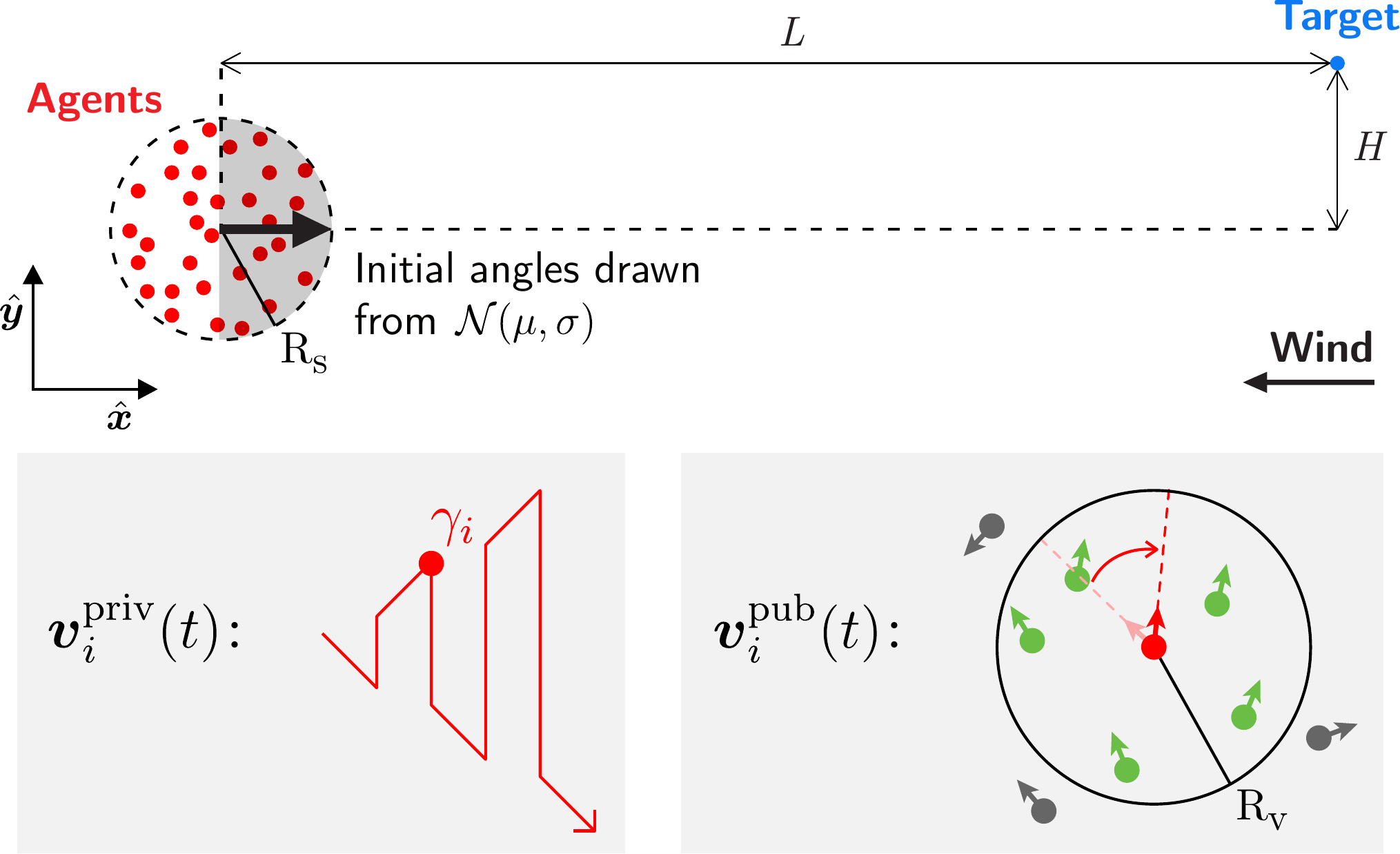}
    \caption{Sketch of the model. (Top) Schematic representation of the geometry of the search task. $\text{R}_s$ denotes the swarm radius while $L$ and $H$ indicate the position of the target relative to the swarm center. The initial heading angles of the agents are drawn from a Gaussian distribution with mean $\mu$ and standard deviation $\sigma$. (Bottom) Illustration of the two behavioral components of motion: the private velocity $\vpriv(t)$ is a cast-and-surge exploratory strategy (left), and the public velocity $\vpub(t)$ results from Vicsek-like alignment with neighbors within the interaction radius $\rv$ (right). At time 0, agents are assigned a random clock $\gamma_i$ with uniform probability between $0$ and $t_{\text{clock}}$, so that $\vpriv(t)=\boldsymbol{v}^{\pm}_{CS}(t+\rc )$, where $\boldsymbol{v}^{+}_{CS}$ represent the cast and surge program depicted in the sketch, and $\boldsymbol{v}^{-}_{CS}$ is its mirror reflected with respect to the x axis. The suffix $\pm$ reflects that each agent follows either the cast and surge as depicted, or its mirror with equal probability.}
    \label{fig:model}
\end{figure}
\section{Model}

 Following Ref.~\cite{Durve2020}, we assume that the velocity of agent $i$ at time $t$, $\vi(t)$, is a combination of the behavioral responses to the private and public cues $\vpriv(t)$ and $\vpub(t)$, prescribed by the following dynamics:
\begin{equation}  \label{eq:main_model}
\begin{cases}
\vpub(t)&=v_0\dfrac{\sum_{j \sim i} \boldsymbol{v}_j(t-\tmem)}{\|\sum_{j \sim i} \boldsymbol{v}_j(t-\tmem) \| } \\
\vi(t) &= v_0 \dfrac{(1-\beta)\vpriv(t) + \beta\vpub(t)}{\| (1-\beta)\vpriv(t) + \beta\vpub(t)\|}
\end{cases}
\end{equation}
where $j\sim i$ represents all neighbors $j$ within an interaction radius from agent $i$. The speed $v_0$ is kept constant throughout the manuscript 
(Table~\ref{tab:parameters}). The search is assumed to be in two dimensions. 

The public velocity $\vpub(t)$ is defined as the average velocity of the neighbors of agent $i$, evaluated at a previous time $t - \tmem$. Neighbors are all agents $j$ located within an interaction radius $\rv$ from agent $i$ at time $t$. This radius may represent a visual radius, or any other mechanism that allows the agents to exchange information, including e.g. neural signaling within an organism with a distributed sensory system. The delay $\tmem$ represents the time required to observe neighbors and process their motion, here modeled as a simple time lag to compute the average direction. While more general formulations using temporal integration kernels are possible, they do not introduce conceptual differences, and we adopt this minimal form for simplicity. As we show below, $\tmem$ plays a central role in shaping the collective dynamics. Note that the cast and surge is directed along the $x$ direction, considered fixed. In the context of olfactory navigation where casting aligns upwind, fixing $x$ amounts to assuming that the agents know the direction the wind blows from.

The private velocity $\vpriv(t)$ is an individual search behavior, here modeled as in Ref.~\cite{Durve2020} by a cast-and-surge strategy widely observed in insects \cite{Vickers2000, David1983, KENNEDY1983, ELKINTON1987} and formalized in \cite{baker_cast_and_surge,shraiman_balkovski}. It produces an anisotropic search trajectory that relies on a clock to alternate between upwind surges and crosswind casts of increasing amplitude, as illustrated in \cref{fig:model}. The clock is reset to zero at each odor detection. In the following, we work in the absence of odor thus we consider no reset: the clock keeps ticking. The model may be extended to include more general private behaviors, such as biased random walks \cite{Codling2008}.

\begin{figure}[ht!]
    \centering
    \includegraphics[width=\columnwidth]{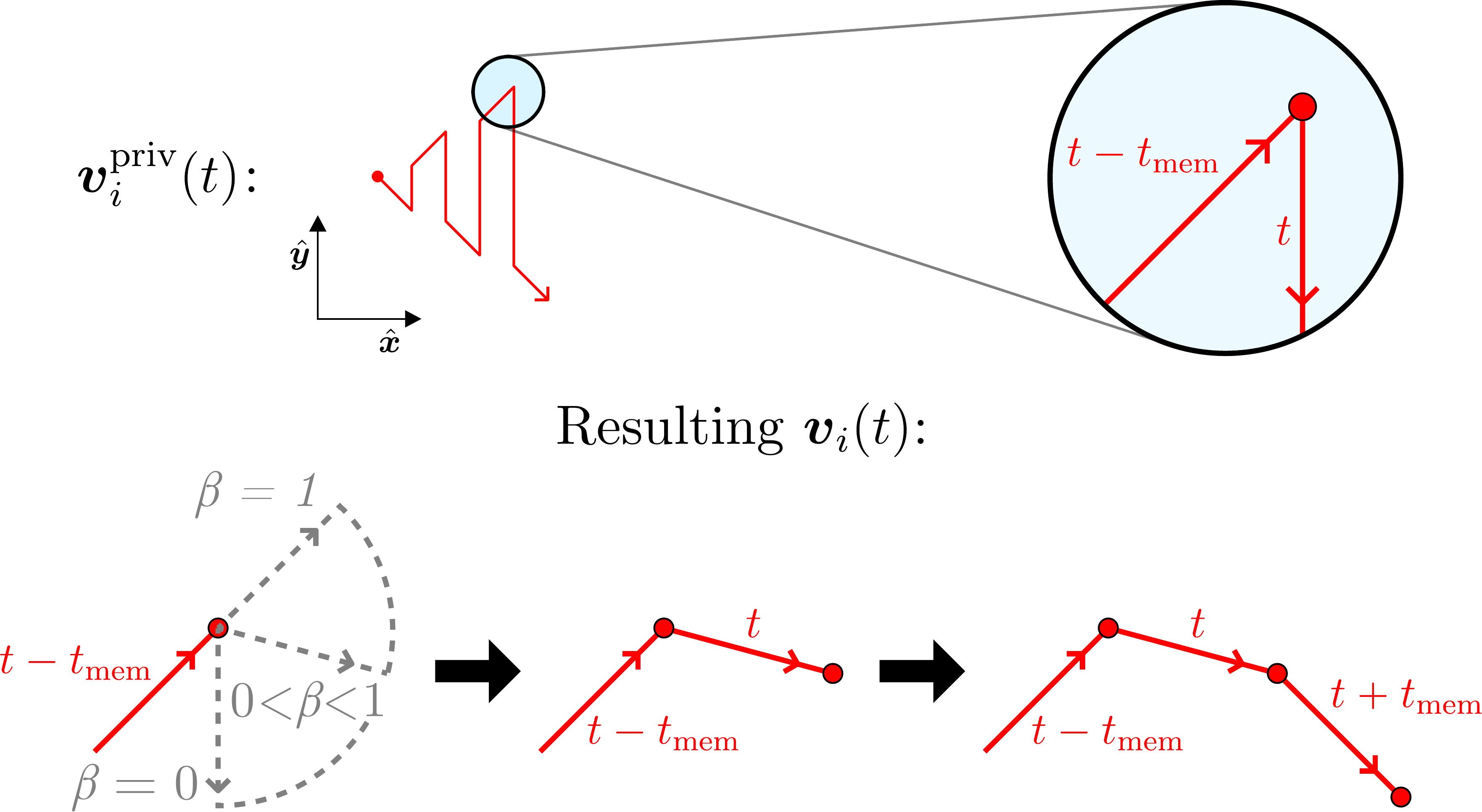}   
    \caption{Sketch illustrating the effect of the trust parameter $\beta$ on the trajectory of a casting agent interacting with another casting agent. Each agent combines its private velocity $\vpriv(t)$, prescribing a sharp turn, with the public velocity $\vpub(t)$, which corresponds to the delayed velocity of its neighbor at time $t-\tmem$. As $\beta$ increases, imitation dominates over sharp turning, resulting in smoothed, more elongated casting trajectories.}
    \label{fig:trajectory_beta}
\end{figure}
\begin{figure}[ht]
    \centering
    \includegraphics[width=.75\columnwidth]{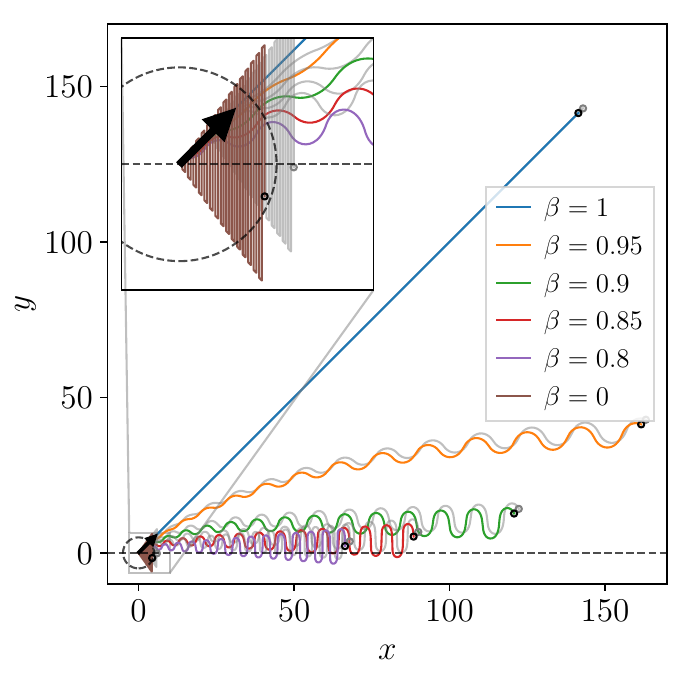}
    \caption{Representative trajectories of pairs of agents for different values of the trust parameter $\beta$ illustrating the effect of social alignment. Solid lines: trajectory of one of the two agents (colors correspond to different values of $\beta$, see legend); gray lines: trajectory of the second agent. Inset: zoom in to visualize $\beta=0$.  
    At $\beta=0$, the dynamics are purely governed by private behavior, resulting in stereotypical casting dynamics. At $\beta=1$, agents follow exclusively public behavior, leading to ballistic motion in the initial heading direction defined by the average initial condition, here oriented at $45\deg$ relative to the centerline, which is marked with a black arrow. For $0<\beta<1$, trajectories smoothly interpolate between the casting and ballistic regimes. Note: for simplicity we choose $\tmem=\delta t$ where $\delta t$ is the numerical time step. However results depend weakly on $\delta t$ as long as $\delta t<\tmem$.}
    
    \label{fig:trust_sketch}
\end{figure}

Before turning to the full simulations of collective search, we illustrate a key  mechanism that links the trust parameter $\beta$ and the integration time $\tmem$ to geometry. Consider a pair of agents moving synchronously with the same initialization, so they follow identical dynamics (in simulations, we introduce a small spatial offset for visualization).
Let us focus on a specific moment $t$ when the private behavior induces a sharp turn from the diagonal heading direction $(\boldsymbol{\hat x}+\boldsymbol{\hat y})\sqrt{2}/2$ to the downward heading direction $-\boldsymbol{\hat y}$, see sketch in \cref{fig:trajectory_beta}. Each of the two agents attempts its abrupt turn, but the actual direction of motion $\vi(t)$ results from the linear combination of $\vpriv(t)=-v_0\boldsymbol{\hat y}$ and the public contribution, which is the velocity of the other agent at time $t-\tmem$ and is still pointing at $45^\circ$: $\vpub(t)=v_0\sqrt{2}(\boldsymbol{\hat x}+\boldsymbol{\hat y})/2$. 
As $\beta$ increases from 0 to 1, the turning angle is also increased, effectively smoothing out the sharp turn prescribed by the private behavior. Because the intensity of the velocity vector is kept constant, an increased trust effectively stretches upwind the agent's casting trajectory. In the extreme case of $\beta=1$, trajectories are ballistic and follow the initial condition without ever aligning to the private behavior. This is confirmed by numerical simulations of $N=2$ interacting agents, for six different values of the trust parameter (see \cref{fig:trust_sketch}).
Both agents eventually turn towards $-\boldsymbol{\hat y}$, after a timescale $\propto \tmem/(1-\beta)$ (see Supplementary Material). Thus $\tmem$ and $\beta$ concur into stretching and smoothing the trajectory of the pair of agents. 

The behavior of the synchronous pair showcases that the strength of imitation of the neighbor coupled to the integration time leads to a delay in turning, i.e. to inertia, as an emerging property of collective behavior. In section IV, we will show that the inertial timescale $\tmem/(1-\beta)$ extends to many agents and defines how fast the center of mass of a group tends to align to the private behavior. This emerging concept of inertia is the fundamental mechanism behind group behavior: the integration time $\tmem$ and the strength of trust $\beta$ dictate the collective time of alignment and therefore the geometry of the area explored by the group, with large $\beta$s favoring slim upwind regions and small $\beta$s favoring exploration. 
Clearly, depending on where the target is relative to the group initial position, more or less exploration is needed. 

\section{Numerical results}

We conduct a series of agent-based numerical simulations, where we follow the trajectories of $N=100$ agents, starting from a random location within a circle of radius $\rs$ centered at the origin of the coordinate system  (see sketch in \cref{fig:model}). At time $t=0$, each agent is headed in a random direction, drawn from a Gaussian distribution with mean $\mu$ and standard deviation $\sigma$ and is assigned a random clock $\gamma_i$ with uniform probability between $0$ and $t_{\text{clock}}$, so that $\vpriv(t)=v_{CS}^{\pm}(t+\rc)$, where $v^{\pm}_{CS}$ is the cast and surge program ($+$) or its mirror reflected with respect to the x axis ($-$) and each agent is assigned the $+$ or $-$ with probability $1/2$ at time $0$. The target is located at coordinates $(L,H)$. Agents orient their private behavior against the mean wind  (direction $\boldsymbol{\hat{x}}$), which is considered known, and align with neighbors within their interaction radius  $\rv$ according to the model \cref{eq:main_model}. For each condition, we conduct $N_S=50$ simulations, and each simulation $r$ stops at time  $T_r$, once it either overshoots the target (failure) or an agent reaches a circle of size $\rd$ around the target (success,  first passage time $T_r$).

To quantify the efficiency of the swarm in reaching a target, we define two complementary performance metrics for the speed and reliability of the search process.
The first is the (empirical) average normalized first passage time (FPT) of all the $S$ successful simulations $\tau=1/S\sum_{r=1}^ST_r/T_{\text{min}}$ normalized by the straight-line time $T_{\text{min}}$ to reach the target. This metric captures how quickly the swarm (or at least one of its members) can locate the target. 
The second is the success rate $\rho=S/N_S$, defined as the fraction of successful simulation trials, in which at least one agent reaches the target within the time horizon. Importantly, $\rho$ is computed across many realizations of the same initial condition, and is not a fraction of agents per simulation, but a fraction of successful trials over the total. Thus $\rho$ reflects the overall reliability of the search strategy.
As we will see next, when agents can break away from the swarm $\tau$ is the most relevant metric of success, as at least one of the agents nearly always reaches the target thus $\rho$ is always $\approx 1$. Conversely, when the group remains cohesive, the success rate $\rho$ is the most relevant metric of success.

\begin{table}[ht]
    \centering
    \caption{Fixed parameters in all simulations.}\label{tab:parameters}
    \begin{tabular}{lccccccc}
        \hline
        $v_0$ & $N$ & $\rv$ & $\rs$ & $L$ & $\tmem$ & $N_S$ \\
        \hline
        0.2 & 100 & 1 & 1 & 75 & 1 & 50 \\
        \hline
    \end{tabular}
\end{table}

\subsection{Finite interaction range}

We begin by analyzing the case where agents interact through a finite interaction range $\rv=1$. Throughout our simulations, the average normalized first-passage time $\tau$ exhibits a minimum at an optimal trust value $\beta^*_\tau$ which depends sensibly on the initial condition and the location of the target. 

\begin{figure}[ht]
    \centering
    \includegraphics[width=\columnwidth]{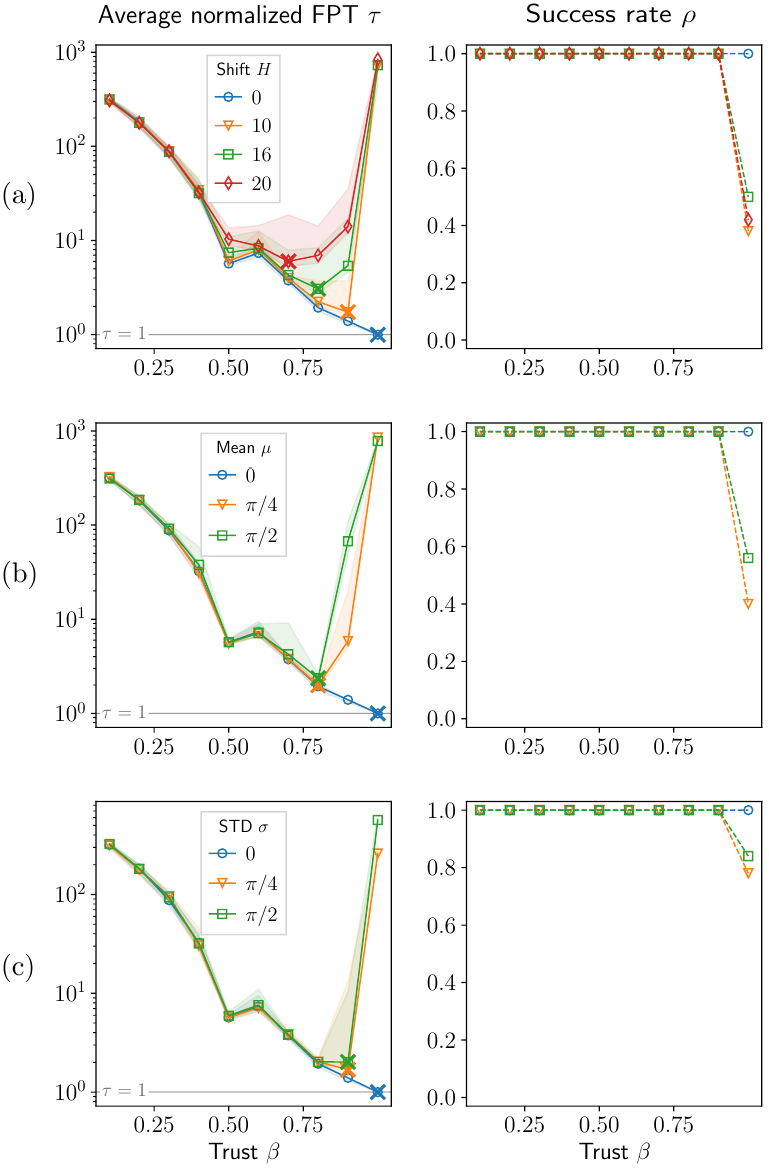}
    \caption{Performance measures for finite interaction range $\rv=1$ under different initial conditions: (a) varying the shift $H$ of the target from the centerline, with $\mu, \sigma = 0$, (b) varying the mean initial angle $\mu$, with $H, \sigma = 0$ and (c) varying the angular dispersion $\sigma$, with $H, \mu = 0$. Left column: normalized first-passage time $\tau$ (averaged over 50 simulations) as a function of $\beta$. Right column: corresponding success rate $\rho$, defined as the fraction of successful trials out of 50 total simulations. Crosses in left column mark $\beta^*_\tau$.}
    \label{fig:perf_finite}
\end{figure}

When all agents are headed upwind at time $0$ ($\mu=\sigma=0$), the optimal trust systematically decreases as $H$ increases, from $\beta^*_\tau \simeq 1$ for $H = 0$ to around $\beta^*_\tau \simeq 0.7$ for $H = 20$. This suggests that, when the target is located further from the midline, agents benefit from reducing imitation in favor of individual exploration (see \cref{fig:perf_finite}~(a)). When agents are headed at an angle relative to the midline at time $0$ and the target is on the midline ($\mu=0,\pi/4,\pi/2$, $\sigma=0$, $H=0$) the optimal trust parameter decreases from $\beta^*_\tau=1$ to $\beta^*_\tau=0.8$ (see \cref{fig:perf_finite}~(b)). And finally, when agents are headed at random angles with $\mu=0$ and $\sigma=0,\pi/4,\pi/2$ and $H=0$, again $\beta^*_\tau$ decreases with further scatter of the initial condition (see \cref{fig:perf_finite}~(c)). All other parameters are kept fixed (see \cref{tab:parameters}). Thus the optimal balance between imitation of others and individual exploration depends on where the target is and where are the agents initially headed. 

In contrast, the success rate $\rho$ remains close to 1 across most values of $\beta$, indicating that the swarm successfully finds the target in nearly every trial, regardless of the trust level (see \cref{fig:perf_finite}, right column). An exception occurs at $\beta = 1$, where $\rho$ drops for all three cases. This drop is due to the total absence of the private contribution: in this limit, agents fully rely on imitation and ignore their individual exploration behavior, moving ballistically in the initial direction set by the group’s configuration at $t = 0$. Unless the swarm aims directly at the target (e.g. for $H=\mu=\sigma=0$), ballistic search is inefficient at finding the target and $\rho$ drops. A nonzero convergence $\rho$ remains because agents with finite-range interactions occasionally detach from the bulk of the swarm and explore independently according to their private dynamics (the cast-and-surge mechanism defined by $\vpriv$ in \cref{eq:main_model}). When this happens, the explorer most likely reaches the target, but takes an extremely long time (hence the extremely high peak in $\tau$). 

\subsection{Infinite interaction range}
\begin{figure}[ht]
    \centering
    \includegraphics[width=\columnwidth]{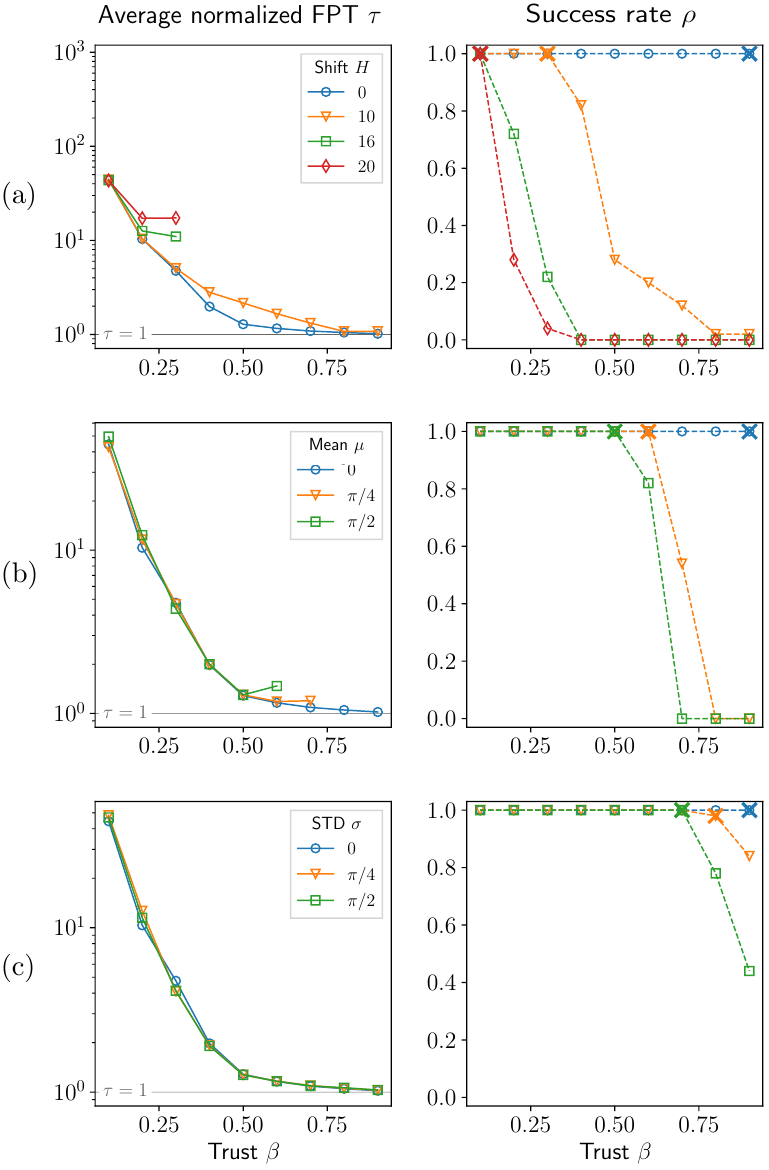}
    \caption{Performance measures for infinite interaction range $\rv$ under the same conditions as in \cref{fig:perf_finite}. 
    Missing points in the $\tau$ plots correspond to cases where $\rho=0$, i.e. no agents reached the target. Crosses in right column mark $\beta^*$.}
    \label{fig:perf_infinite}
\end{figure}

We now, and for the rest of the paper, focus on the case where all agents interact with all others at all times, $\rv\rightarrow \infty$. In this case, agents cannot detach from the group and all  contribute and respond to the same social signal, akin to cells belonging to the same organism or robotic swarms with global communication. 
\begin{figure*}[ht]
    \includegraphics[width=0.9\textwidth]{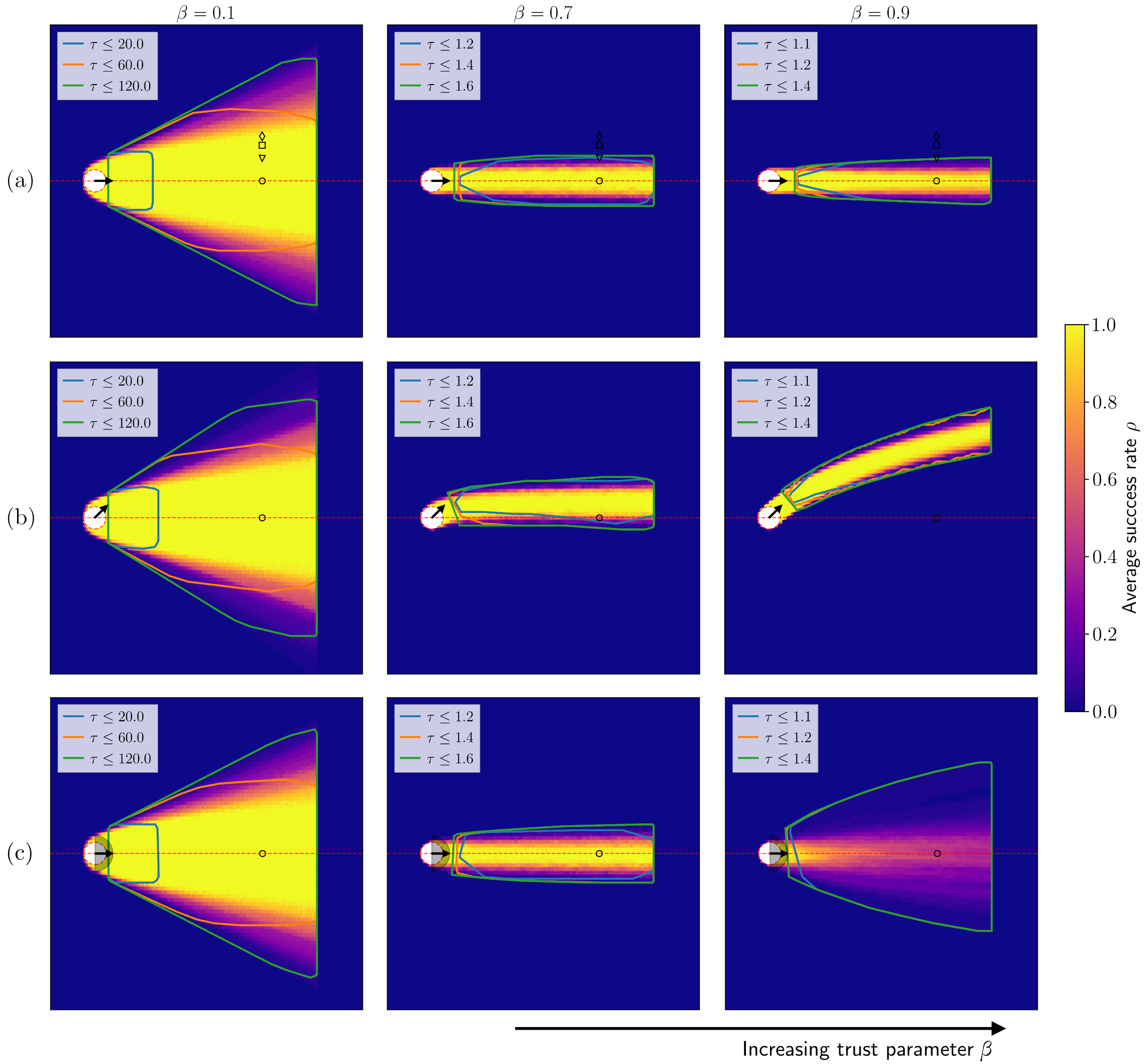}
    \caption{Spatial performance maps for three values of the trust parameter $\beta$ (columns). Rows correspond to different initial conditions: (a) $\mu=0$, $\sigma=0$; (b) $\mu=\pi/4$, $\sigma=0$; (c) $\mu=0$, $\sigma=\pi/2$. The colormap represents the average success rate $\rho$ over 50 independent simulations. Contours indicate regions where the normalized first-passage time $\tau$ falls below selected thresholds (see legends), providing a spatial visualization of search efficiency across space.
}
    \label{fig:maps}
\end{figure*}
Because no agent can break away from the group, the outcome is binary: either the entire swarm succeeds, or it collectively fails. As a result, $\rho$ now decreases dramatically  from nearly $100\%$ at $\beta=0$ and drops to near zero at finite $\beta$, making $\rho$ a much more sensitive indicator of group performance than $\tau$ (see \cref{fig:perf_infinite}, right column). Thus, here, an optimal balance between private and social cues emerges to ensure consistent -- rather than quick -- target location, and we call this optimal trust parameter $\beta^*$ to distinguish it from $\beta^*_\tau$, which minimizes time. Again, the optimal trust parameter, $\beta^*$, depends sensibly on the initial heading of the agents, as well as on target location. Note, however, that to ensure maximum convergence, $\beta^*$ needs to be pushed to much lower values than for the finite interaction (compare star in \cref{fig:perf_finite}, left column and \cref{fig:perf_infinite}, right column). Thus, individual exploration needs to be weighed in heavily to prevent the swarm from missing the target entirely. As a result, all agents reach the target, but when the target is off the centerline, navigation is much slower than with finite-range interaction (compare $\tau(\beta^*)$ in \cref{fig:perf_infinite}, top row and $\tau(\beta^*_\tau)$ in \cref{fig:perf_finite}, top row).

\begin{figure*}[ht]
    \centering
    \includegraphics[width=0.9\textwidth]{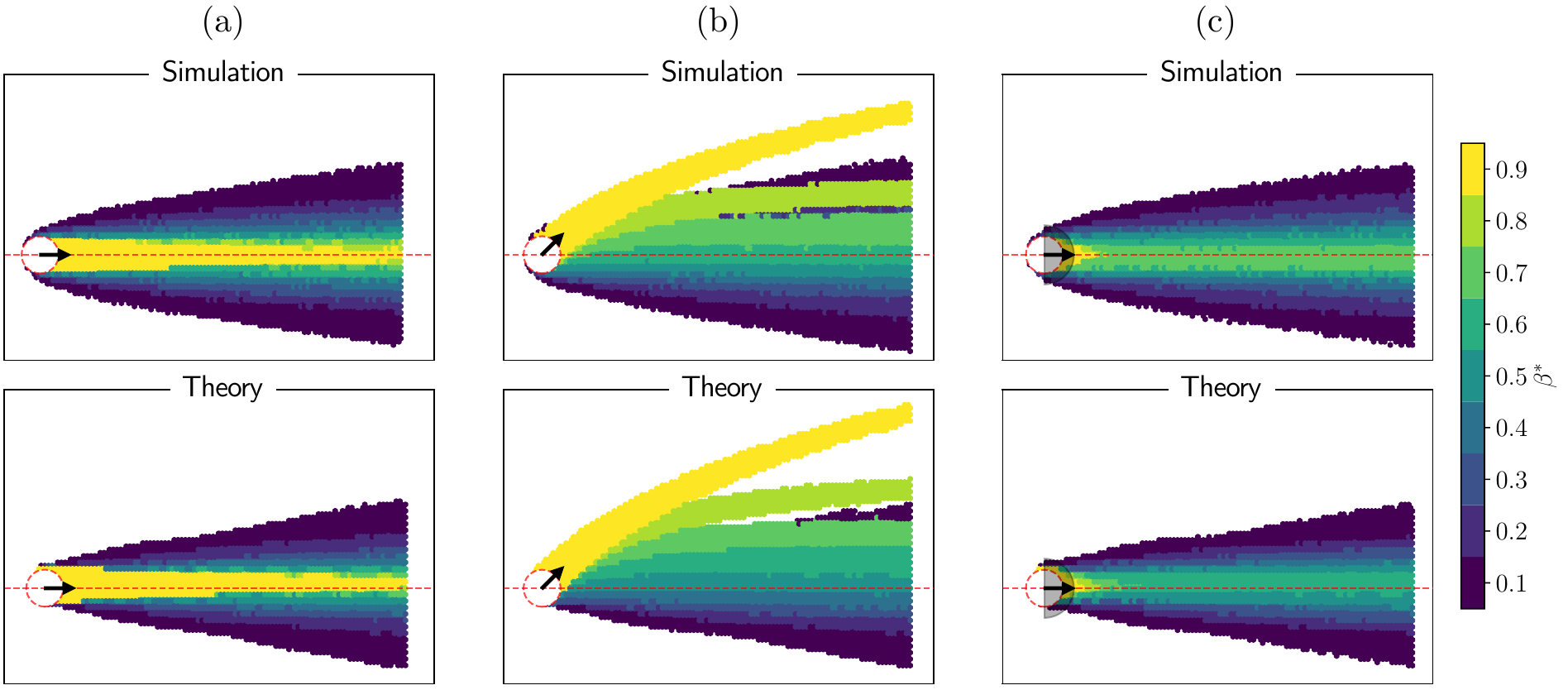}      
    \caption{Comparison between simulations (top row) and theoretical predictions (bottom row) for the optimal trust parameter $\beta^*$, across different initial conditions: (a) $\mu=0$, $\sigma=0$; (b) $\mu=\pi/4$, $\sigma=0$; (c) $\mu=0$, $\sigma=\pi/2$ (as in \Cref{fig:optimal_maps}). The theoretical maps are obtained by numerically inverting \cref{eq:rho} for each $(L,H)$ pair, using the center-of-mass trajectory predicted by \cref{eq:cm_betaalti} and \cref{eq:cm_betabassi} and $|v_{cm}(t)| = \beta v_0$.}    
    \label{fig:optimal_maps}
\end{figure*}

\subsection{Spatial maps of performance and optimality}

Next, we quantify how the swarm explores space, by performing many simulations where we systematically vary target location within a regular grid and visualize maps of performance. 
At $\beta = 0.1$ (see \cref{fig:maps}, leftmost column), agents explore space widely, leading to an extended region where $\rho$ is close to 1. However, they are also very slow, as indicated by the $\tau$ contours. At such low trust, the initial condition plays a minor role: agents quickly ``forget" their initial direction, and the exploration pattern looks very similar across all three initializations (rows matching initial conditions in rows of \cref{fig:perf_infinite}).
At higher values of $\beta$ (second and third columns), exploration becomes much more directional, and the shape of the explored region depends sensibly on the initial condition. Note that, somewhat unintuitively, when initial angles are drawn from a Gaussian distribution with nonzero standard deviation (third row, with $\sigma = \pi/4$), we observe a broadening of the explored region at large $\beta$. However, this broader coverage comes with lower average success rates: each individual trajectory remains highly directional (as in the first two rows), and only their ensemble average produces a wider exploration area.

These maps clearly illustrate that the optimal balance between individual and collective signals varies dramatically in space. Scanning through all values of $\beta$ for each target location, we define the optimal trust parameter to ensure reliable target localization ($\beta^*$ is the largest value of $\beta$, conditioned to  $\rho\ge 0.95$). Maps of $\beta^*$ (see \cref{fig:optimal_maps}, top row) reveal that the optimal trust parameter depends on both target location and heading angle of the agents at time $0$ (see columns in \cref{fig:optimal_maps} matching initial conditions in rows of \cref{fig:perf_infinite}). 
An alternative optimal trust $\beta^*_\tau$ can be defined by forfeiting reliable target location and minimizing the time to reach the target, regardless of the success rate. As already observed before, this definition of optimality may be less significant for the infinite $\rv$, as it ends up selecting much larger values of $\beta^*$, at which convergence is poor (see Supplementary Fig.~S1).

\section{Analytical results}

To predict how the optimal trust depends on the parameters of the model, we develop an asymptotic theory that is valid in the limit of large number of agents $N$ and infinite interaction radius. We summarize the main results here and refer the interested reader to the Supplementary Material for the analytical derivation and extensions to more general forms of private behaviors. The starting point for the calculation is that for large $N$, the public velocity is aligned with the velocity of the center of mass of the swarm, $\vcm$, 
\begin{equation}
\label{eq:vpub}
\vpub(t) \approx v_0 \frac{ \vcm ( t - \tmem)}{\|\vcm( t-\tmem)\|}
\end{equation} 
 Therefore, as long as $v_0/N \ll \| \vcm\|$, we can assume that $\vpub$ does not depend on $i$.

\subsection{High trust parameter $\beta\rightarrow 1$: Langevin dynamics}

First, by expanding the nonlinearity in \cref{eq:main_model}, we find that agents are mostly aligned to the public velocity with a perturbation originating from their private exploration:
\begin{equation}
\label{eq:proj}
\vi (t) \approx \vpub(t) + (1-\beta) P_{\vpub(t)^{\perp}} (\vpriv(t))
\end{equation}
 where $P_v: \mathbb{R}^2 \to \mathbb{R}^2$ is the projection operator on the linear space generated by $v$, that is $P_v(w)= \frac{\langle v, w\rangle}{\|v\|^2} v$. Using \cref{eq:vpub} and noting that $\|\vcm(t)\| \approx v_0$ to within $O((1-\beta)^2)$, we obtain a recursive equation for $\vi (t)$ that can be interpreted as the discretization of the following ODE: 
\begin{equation}
 \label{eq:langevin}
\dot{\theta}(t)= \frac {(1-\beta)w(t)}{\tmem}  \cdot \sin( \theta_w(t)- \theta(t))
\end{equation}
 where $w(t)$ is the intensity of the mean private velocity and $\theta$ and $\theta_w$ are the heading directions of the center of mass and of the mean private velocity respectively, i.e. $\vcm(t)=v_0 e^{i \theta(t)}$ and $\langle \vpriv(t) \rangle = w(t) e^{i \theta_w(t)}$. We integrate the Langevin equation \cref{eq:langevin}, starting from initial conditions for the two heading angles $\theta_w(t=0)$ and $\theta(t=0)$ computed from the empirical average of the heading angles of the agents at time 0. Clearly, these empirical averages depend on the parameters $\mu$ and $\sigma$ that define the initialization of the agents. Specifically, for $N\rightarrow\infty$, $\theta_w(t=0)\rightarrow\mu$  and $\theta(t=0)\rightarrow\mu$, while $w(t=0)\rightarrow e^{-\sigma^2}$.
\cref{eq:langevin} dictates that the center of mass $\vcm$ aligns to the expected private velocity, $\langle \vpriv(t) \rangle$, with a
timescale $\tmem/(1-\beta)$: this timescale defines the collective mechanism of inertia introduced above and emerges from a compromise between imitation and private behavior. At larger $\beta$, when agents imitate their neighbors more,  the timescale $\tmem/(1-\beta)$ grows: the group align  slowly to the expected private behavior and travels for longer on a nearly ballistic trajectory along a direction selected by the initial condition. \cref{eq:langevin} can be integrated to obtain the trajectory of the center of mass
$(x_{cm}(t),y_{cm}(t))$:
\begin{equation}
\label{eq:cm_betaalti}
(x_{cm}(t),y_{cm}(t)) = \int_0^t|\vcm(s)|(\cos \theta(s),\sin \theta(s))ds\end{equation}
 where $|\vcm(s)|=\beta v_0$, with initial condition $(x_{cm}(0),y_{cm}(0))=(0,0)$. 

Next, we analyze how agents spread around the center of mass by focusing on $\sigma_y(t)= \langle (y_i(t) - y_{cm}(t))^2 \rangle_i$, where $\langle \cdot \rangle_i$ is the expected value over agents. Using \cref{eq:proj} and given that the cast-and-surge private behavior is mostly directed crosswind, we can show that:
\begin{equation}
\sigma^2_y(t) = \sigma^2_y(0)+ \frac{(1-\beta)^2 c_t^2}{t_{\text{clock}}} \int_0^{t_{\text{clock}}} (y^{\text{CS}}(t+\gamma)- y^{\text{CS}}(\gamma))^2 \, d\gamma
\label{eq:sigmay_betaalti}
\end{equation}
where $\gamma$ is initialized randomly between $0$ and ${t_{\text{clock}}}$ for each agent, as described above (see caption of \cref{fig:model}). Here we have substituted the expected value over agent trajectories as an integral over the initial clock $\gamma$ of the cast-and-surge program; $y^{\text{CS}}$ indicates the cast-and-surge program starting from the beginning i.e. clock $\gamma=0$ (entire trajectory in \cref{fig:model}, bottom left). 
The constant $c_t$ represents the alignment of the center of mass in the upwind direction  ($c_t=1$ when the center of mass moves upwind, $c_t=0$ if the center of mass moves crosswind). Note that the swarm spreads little from its initial condition $\sigma_y(0)$, as the second term in \cref{eq:sigmay_betaalti} depends on $(1-\beta)^2$. Because the second term is small, we ignore corrections to $c_t=1$ which would be at play for initial conditions different from $(\mu,\sigma)=(0,0)$. 
Importantly, the integral in \cref{eq:sigmay_betaalti} can be calculated explicitly knowing the private velocity, as it does not depend on the kinematics of the center of mass.

\subsection{Low trust parameter $\beta\rightarrow 0$}

Starting as before from \cref{eq:vpub} and noting that the probability that an agent's private velocity heads upwind is small, $O(1/\sqrt{N})$, we can compute the averages analytically. The effect of the nonlinearity in \cref{eq:main_model} is that the swarm quickly orients upwind, regardless of the initial conditions, with a small additive noise: 
\begin{equation}
\label{eq:vcm_smallbeta}\vcm(t)=\beta (v_0,0) + \boldsymbol{\eta}(t)+ O(\beta^2),
\end{equation}
 where the noise $\boldsymbol{\eta}(t) \approx(1-\beta)(1+\beta(1-\beta)))\langle \vpriv (t)\rangle\sim O(v_0/\sqrt{N})$. \cref{eq:vcm_smallbeta} can be integrated to yield a prediction for the trajectory of the center of mass:
\begin{equation}
\label{eq:cm_betabassi}
x_{cm}(t)= \beta v_0 t ;\quad y_{cm}(t)= (1-\beta)(1+\beta(1-\beta)) \langle y_i^{CS}(t)\rangle
\end{equation}
 where the $\langle\cdot\rangle$ is the empirical average over the random clock $\gamma$ and orientation of the agents.
We calculate agents' deviation from the trajectory of the center of mass to obtain the spread of the swarm:
\begin{align}
\sigma^2_y(t) = \sigma^2_y(0)+ [1+(2\beta(1-\beta))]\frac{(1-\beta)^2}{t_{\text{clock}}} \times \nonumber \\
\times\int_0^{t_{\text{clock}}}(y^{\text{CS}}(t+\gamma)- y^{\text{CS}}(\gamma))^2 \, d\gamma
\label{eq:sigmay_betabassi}
\end{align}

Eqs.~\eqref{eq:cm_betaalti} and \eqref{eq:sigmay_betaalti} (for large $\beta$) and Eqs.~\eqref{eq:cm_betabassi} and \eqref{eq:sigmay_betabassi} (for small $\beta$) describe asymptotically the trajectory and extension of the swarm in time and serve as a basis to predict the probability to reach a target at any location in space, as elaborated below.  

\subsection{Comparison between theoretical predictions and simulations}

We now derive analytical predictions for the success rate $\rho$ throughout space as a function of $\beta$. We model the swarm as a Gaussian cloud centered around the $(x_{cm},y_{cm})$, according to Eqs.~\eqref{eq:cm_betaalti} and \eqref{eq:cm_betabassi} for large and small $\beta$ respectively. The horizontal size of the swarm remains stable around the initial swarm size, $\rs$, as 
agents move upwind with a rather similar speed, with variations only in correspondence with the diagonal step in the private program, which occur rarely. 
The vertical extent of the swarm is quantified by the standard deviation of the agent's $y$-coordinates around the center of mass, $\sigma_y$. The two asymptotic expansions at large and small $\beta$, Eqs.~\eqref{eq:sigmay_betaalti} and \eqref{eq:sigmay_betabassi}, can be combined into a single expression that is valid across all values of $\beta$: 
\begin{align}\nonumber
    \sigma_y(t)^2 &\simeq \left(\frac{\text{R}_\text{s}}{2}\right)^2 + \left[ 1+2 \beta(1-\beta) \right]\frac{(1-\beta)^2}{t_\text{clock}}\times\\
    &\times\int_0^{t_{\text{clock}}}(y^{\text{CS}}(t+\gamma)- y^{\text{CS}}(\gamma))^2 \, d\gamma \label{eq:sigmay}
\end{align}
which interpolates the two expressions and uses $\sigma_y(t=0)=\rs/2$ where $\rs$ is the radius of the swarm. 
Matching of these two asymptotic expansions is sufficient to explain the results, and no further expansion is needed around $\beta=0.5$.
Similarly, although in principle leading order corrections distinguish \cref{eq:cm_betaalti} and \cref{eq:cm_betabassi}, we use \cref{eq:cm_betaalti} for all values of $\beta$ suggesting corrections remain numerically small. 

To produce a theoretical map of the optimal $\beta^*$ in space, we obtain 50 trajectories of the theoretical swarm for 10 values of $\beta\in[0,1]$. For each trajectory, we compute the empirical average that dictates the initial condition $\theta(t=0)$ for the center of mass and for the expected private behavior $\theta_w(t=0)$ and we trace the trajectory of the center of mass by integrating \cref{eq:cm_betaalti}. 

At each time, the swarm is modeled as an ellipse, whose horizontal semiaxis is $\rs$ 
and vertical semiaxis is $2\sigma_y$ obtained from \cref{eq:sigmay}. A target at a location $(L,H)$ is declared reached by a swarm when any point within the ellipse overlaps with a circle of radius $\rd$ centered at $(L,H)$. Scanning through all values of $\beta$, we identify $\beta^*$ as the largest  value of $\beta$ that guarantees $95\%$ of the realizations reach the target at location $(L,H)$. The theoretical maps of $\beta^*$ thus obtained are represented in the bottom line of \cref{fig:optimal_maps}, and are in very good agreement with the agent based simulations for all initial conditions.

We now seek to derive an analytical prediction for the performance as a function of $\beta$, rather than focusing only on $\beta^*$. For this section, we focus on the initial condition $\mu=\sigma=0$, for targets off centerline, as in \cref{fig:perf_infinite}~(a). We start by observing that, mathematically, for the search to be successful, the target must be within the region occupied by the swarm. Because the swarm needs to cover a horizontal distance $L$, its  horizontal speed is $\approx\beta v_0$ and its $x$-width is $\rs$, the target may only be reached within an admissible time window $t\in[t_-,t_+]$ with $t_\pm=\frac{L\pm R_\text{s}}{\beta v_0}$.
Within the admissible time window, the swarm's maximum $y$-extent is:
\begin{equation}
    \hat{\sigma}_y = \max_{t\in[t_-,t_+]
    } \sigma_y(t) \quad\text{ and } \quad\hat{t}=\argmax_{t\in[t_-,t_+]} \sigma_y(t)
\end{equation}

Qualitatively, we expect that if $\hat\sigma_y< H$ then no agent is able to reach the target and $\rho\approx 0$, whereas if $\hat\sigma_y\gtrsim H$, in the large $N$ limit of dense swarm, $\rho\approx 1$. When $\hat\sigma_y \gg H$, performance will eventually decrease as the swarm widens and dilutes. This effect can be further quantified and will become visible at small $N$, which are not considered here.

\begin{figure}[ht]
    \includegraphics[width=.75\columnwidth]{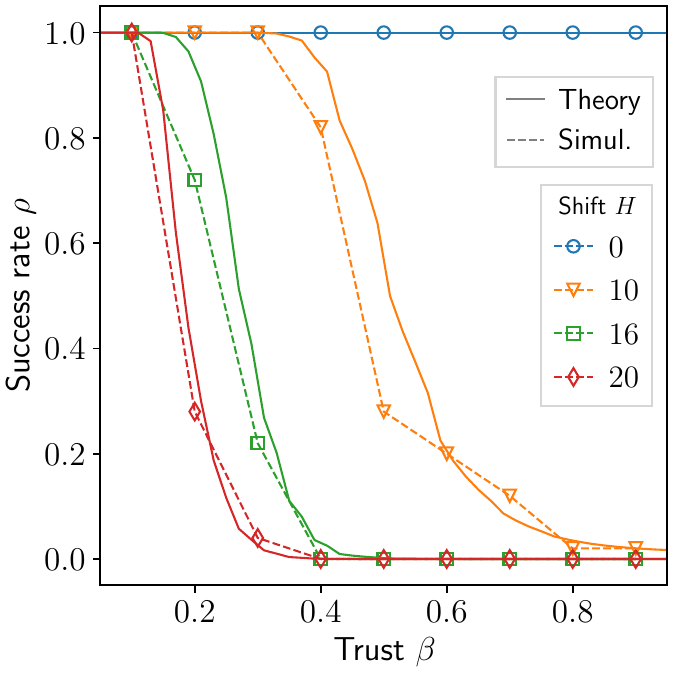}
    \caption{Comparison between numerical simulations (dashed lines) and theoretical predictions from \cref{eq:rho}  (solid lines) for the success rate $\rho$ as a function of the trust parameter $\beta$.  
    Simulations correspond to the same data as in \cref{fig:perf_infinite}~(a).
}
    \label{fig:comparison_straight_theory_rho}
\end{figure}

To predict analytically the transition between $\rho=1$ and $\rho=0$, we consider the distribution of agents around the center of mass. We assume that the agent's position $y$ is distributed as a Gaussian centered around the center of mass, with standard deviation $\hat{\sigma}_y$ with cumulative distribution: 

\begin{equation}
F(y)=\frac{1}{2}\left[1-\text{erf}\left(\frac{y}{\hat{\sigma_y}}\right)\right]
\end{equation}
where we neglected the $y$-oscillations of the center of mass around its initial position,
$0$, which are indeed small for the initial condition $\mu=\sigma=0$ considered here. The probability that at least one out of $N$ agents reaches a position $y$ within a distance $\Delta H$ from the target, located at 
$y_{\text{target}}=H$
at a specific time $\hat{t}$ is:
\begin{equation}
    \rho(\beta) = 1- \left[1 - \left(F_{H + \Delta H} - F_{H - \Delta H} \right) \right]^N \quad 
    \label{eq:rho}
\end{equation}
with an implicit  dependence on $\beta$ within $\hat{\sigma}_y$ and $\Delta H$. To complete the argument, we need to quantify $\Delta H$, i.e. the precision with which an agent needs to reach the target,  as dictated by the 
detection radius $\rd$. Each agent spends a time $\Delta t \approx \rd / (\beta v_0)$ with its $x$ position within a distance $\rd$ from the target, as its horizontal speed is $\approx \beta v_0$. Within $\Delta t$ 
its $y$ position moves of $\tilde{y}\approx \sqrt{1-\beta^2}\,v_0\Delta t$. Hence at a specific time, its $y$ position needs to be within an extended distance $\Delta H = \rd+\tilde{y}\approx \rd/\beta$. \cref{eq:rho} with $\Delta H=\frac{\rd}{\beta}$ provides an asymptotic prediction for how the success rate $\rho$ depends on $\beta$ which successfully explains the simulations, see \cref{fig:comparison_straight_theory_rho}. 

\section{Discussion}

Our results show that the optimal balance between individual and social behavior in this model of collective search is strongly constrained by geometry. In particular, the initial heading of the agents and the relative position of the target determine whether imitation (high trust) or individuality (low trust) leads to better performance. Note that, although our model does not include odor cues, our results are qualitatively consistent with those obtained in olfactory search scenarios \cite{Durve2020}, suggesting that geometry may remain relevant even in a chemically structured environment. 
Specifically, the simulations that are most closely related to Ref.~\cite{Durve2020} are those with finite interaction radius $\rv$ the target in the center line, $H=0$, and non zero average and standard deviation of the initial heading angles, $\sigma$ and $\mu$. From \cref{fig:perf_finite}~(b) and (c) (green and orange curves), we find an optimal $\beta^*_\tau$ between 0.8 and 0.9, consistent with the value reported in Ref.~\cite{Durve2020}. 
Note that in this model, trust dictates to what extent public information affects the private behavior both because public information is shared and also because imitating others affects an agent's trajectory.  
A more complex model where agents can dynamically adapt their level of trust, or selectively weigh different sources of information depending on context will be helpful to disentangle the role of geometry from the role of sharing of sensory information. %
For example, one may replace the linear combination of private and public velocities of the model with a more flexible mechanism, such as passing them as inputs to a neural network trained to optimize search performance. This approach has been considered for the navigation of single microswimmers in complex flows \cite{Mecanna2025}.

The geometric origin of the optimal trust relies on group inertia, which in turn emerges by imitation with a delay, $\tmem$. 
Group inertia has been noted in previous works where alignment combined with a mechanism that preserves past velocities or orientations, can introduce an effective timescale, resulting in smoother, more persistent trajectories \cite{DeKarmakar2022,Benedetto2020,Cavagna2014}. Similarly, we find that when agents align more strongly with their neighbors, they become less reactive to abrupt changes in their private behavior, and follow a smoother trajectory dominated by past social input. The resulting stretching of the trajectories shrinks the area explored by the swarm and ultimately dictates the optimal $\beta^*$.
Low trust enhances wide spatial exploration, while high trust correspond to straighter trajectories and increases the chances of reaching the target quickly, once the direction is correct. 
Optimal performance is achieved when this smoothing effect allows just enough exploration to reach the target.
Clearly, in realistic scenarios the target location is unknown, hence the optimal trust parameter $\beta^*$ cannot be set a priori. Instead, a real swarm may be optimized to reach targets at several potential locations. This requires prior information regarding the region where the target may be located, which may be derived e.g. from prior experience or from sensory information. In the presence of odor, the group may base their search on a model for how wide the odor spreads from a concentrated source. This is the usual approach in Bayesian algorithms for collective olfactory search that use a model for the likelihood of encountering an odor signal in space (see e.g. \cite{masson2009chasing,karpas2017information,panizon2023seeking,hajieghrary2016multi,park2020cooperative}). 

Beyond the context of genuinely multi-agent systems, our results may
offer insights into individual organisms with distributed sensing architectures. Indeed, when all agents interact with all others, the result is that the whole group remains somewhat cohesive. 
Because smaller groups cannot detach from the rest, all agents reach the target or none. 
Interestingly, the cohesion of the group slows it down considerably when the target is off the centerline, as seen by comparing \cref{fig:perf_infinite}~(a) and \cref{fig:perf_finite}~(a). For example, to ensure that all simulations are successful starting from $H=20$, the trust parameter is constrained to be $\beta\le 0.2$ for $\rv\rightarrow\infty$; within this range of $\beta$, the shortest time is $\tau(0.2)\approx 45$ (red curve). In contrast, for finite interaction radius $\rv=1$, all simulations are successful as long as $\beta\le0.85$ and within this range the shortest time is $\tau(0.7)\approx 6$, thus when the swarm is allowed to break into smaller and faster subgroups, the subgroups can reach the target almost 8 times as fast as the cohesive group. Further work is needed to establish how information should be shared to speed up the search as a function of how cohesive the group is.

Octopus navigation with a flexible body plan that actively switches from compact to extended, 
represents a fascinating example of collective search \cite{weertman2024octopus}. 
Sensory-motor control is partly decentralized, with peripheral ganglia operating semi-independently from the central brain \cite{Zullo2011, Sumbre2001, GodfreySmith2016}. 
How ``private'' input  from local sensing at the periphery is combined with ``public'' descending commands from the central nervous system is a fascinating open question of current research. Our results suggest that geometry may play a role in setting the balance between local and central commands and may be extended to establish how this balance shifts in response to the remarkable changes in shape of these fascinating soft bodied animals.  

\begin{acknowledgments}
This research was supported by grants to AS from the European Research Council (ERC) under the European Union’s Horizon 2020 research and innovation programme (grant agreement No 101002724 RIDING) and the National Institutes of Health (NIH) under award number R01DC018789. The European Commission and the other organizations are not responsible for any use that may be made of the information it contains.
\end{acknowledgments}

\bibliography{ref}

\end{document}